\documentstyle[11pt]{article}

\makeatletter
\@addtoreset{equation}{section}
\makeatother

\def\dalemb#1#2{{\vbox{\hrule height .#2pt
        \hbox{\vrule width.#2pt height#1pt \kern#1pt
                \vrule width.#2pt}
        \hrule height.#2pt}}}

\let\a=\alpha \let\b=\beta   \let\e=\epsilon

 \def\bd{\begin{document}} \def\ed{\end{document}}
\def\ds{\documentstyle} \let\fr=\frac \let\bl=\bigl \let\br=\bigr
\let\Br=\Bigr \let\Bl=\Bigl 
\let\bm=\bibitem
\let\na=\nabla
\let\pa=\partial \let\ov=\overline 
\newcommand{\be}{\begin{equation}} 
\newcommand{\ee}{\end{equation}} 
\def\ve{\varepsilon}
\def\ba{\begin{array}}
\def\ea{\end{array}}
\def\ft#1#2{{\textstyle{{\scriptstyle #1}\over {\scriptstyle #2}}}}
\def\fft#1#2{{#1 \over #2}}
\def\del{\partial}
\def\sst#1{{\scriptscriptstyle #1}}
\def\oneone{\rlap 1\mkern4mu{\rm l}}
\def\e7{E_{7(+7)}}
\def\td{\tilde}
\def\bog{Bogomol'nyi\ }
\newcommand{\ho}[1]{$\, ^{#1}$}
\newcommand{\hoch}[1]{$\, ^{#1}$}
\newcommand{\bea}{\begin{eqnarray}} 
\newcommand{\eea}{\end{eqnarray}} 
\newcommand{\ra}{\rightarrow}
\newcommand{\lra}{\longrightarrow}
\newcommand{\Lra}{\Leftrightarrow}
\newcommand{\ap}{\alpha^\prime}
\newcommand{\bp}{\tilde \beta^\prime}
\newcommand{\tr}{{\rm tr} }
\newcommand{\Tr}{{\rm Tr} } 
\newcommand{\NP}{Nucl. Phys. }
\newcommand{\tamphys}{\it Center for Theoretical Physics,
Texas A\&M University, College Station, Texas 77843}

\newcommand{\auth}{H. L\"u and C.N. Pope}

\begin{document}
\begin{flushright}
\hfill{CTP TAMU-24/96}\\
\hfill{SISSA 103/96/EP}\\
\hfill{hep-th/9607027}\\
\end{flushright}

\vspace{20pt}

\begin{center}
{\large {\bf $SL(N+1,R)$ Toda Solitons in Supergravities }}

\vspace{30pt}

\auth

\vspace{15pt}

{\tamphys}


\vspace{15pt}

{\it SISSA, Via Beirut No. 2-4, 34013 Trieste, Italy }



\vspace{40pt}

\underline{ABSTRACT}
\end{center}

       We construct $(D-3)$-brane and instanton solutions using 
$N \le 10-D$ one-form field strengths in $D$ dimensions, and show
that the equations of motion can be cast into the form of the $SL(N+1,R)$ 
Toda equations.  For generic values of the charges, the solutions are 
non-supersymmetric; however, they reduce to the previously-known 
multiply-charged supersymmetric solutions when appropriate charges vanish.  

{\vfill\leftline{}\vfill
\vskip	10pt
\footnoterule
{\footnotesize
	Research supported in part by DOE 
Grant DE-FG05-91-ER40633 and \vskip	-12pt}  \vskip	10pt
{\footnotesize 
      EC Human Capital and Mobility Programme under contract ERBCHBGCT920176.  
} 
}

\pagebreak
\setcounter{page}{1}

\section{Introduction}

     A convenient procedure for constructing $p$-brane solitons in string
theory or M-theory is first to perform a consistent truncation of the bosonic
sector to a subset of the fields that includes the metric, the dilatonic
scalars $\vec\phi$ and the $n$-index field strengths $F_\a$ that are 
involved in the solution.  In general, we shall concentrate on those theories
that are obtained by dimensional reduction of M-theory.  The $D$-dimensional
bosonic Lagrangian takes the form
\be
e^{-1} {\cal L} = R - \ft12 (\del \vec \phi)^2 - \ft12 \sum_\a
e^{- \vec c_\a \cdot \vec \phi}\, F^2_\a + {\cal L}_{FFA}\ ,\label{genlag0}
\ee
where the ``dilaton vectors'' $\vec c_\a$ are constant vectors, characteristic
of the dimension $D$ and of the field strengths involved.  
Their detailed forms can be found in \cite{lpsol}.\footnote{We shall follow 
the notation of \cite{lpsol}, in which internal compactified indices are 
denoted by 
$i,j,\ldots$, running over $11-D$ values. Thus in $D$ dimensions there is the
4-form $F_4$, and 3-forms $F_3^{(i)}$, 2-forms $F^{(ij)}_2$ and 1-forms
$F_1^{(ijk)}$ coming from the 4-form in $D=11$, and 2-forms ${\cal F}_2^{(i)}$
and 1-forms ${\cal F}_1^{(ij)}$ coming from the $D=11$ metric.  The 1-forms
${\cal F}_1^{(ij)}$, coming from the further dimensional reduction of the
2-forms ${\cal F}_2^{(i)}$ in higher dimensions, are defined only for $j>i$.}
There are also terms, denoted by ${\cal L}_{FFA}$ in 
(\ref{genlag0}), originating from the $F\wedge F\wedge A$ term for the 4-form
field strength in $D=11$.  A further complication is that the $n$-form  
field strengths $F_\a$ are not in general given simply by the exterior
derivatives of $(n-1)$-form potentials; there are additional Chern-Simons 
modifications that result from the dimensional reduction process.  Thus in
general lower-degree potentials also contribute to the $n$-form field 
strengths, and conversely, the $(n-1)$-form potentials can contribute to
fields strengths of degrees higher than $n$.  It is therefore by no means 
automatic that one can perform a consistent truncation to a sector containing 
just the metric, dilatonic scalars $\vec \phi$, and $n$-form field strengths
$F_\a$ that are expressed simply as exterior derivatives of potentials.  In
particular, it should be emphasised that consistency implies
that the vanishing of the truncated fields must be consistent with their own
equations of motion, and that it is not sufficient that they simply be 
absent from the Lagrangian. 

     There are two reasons why it is nonetheless preferable to try to work
only with sets of fields for which the above consistent truncation can be
performed.  The first is that it is much simpler to solve the equations when
the Chern-Simons and $FFA$ terms are not active.  The second reason is that
in fact many of the solutions where the Chern-Simons and $FFA$ terms
play r\^ole are nothing but U-duality rotations of simpler solutions where
the Chern-Simons and $FFA$ terms are not active.  Thus a convenient 
strategy for enumerating
solutions is first to look for those where the Chern-Simons and $FFA$ terms
do not contribute, and then, if desired, to act on these with U-duality in
order to fill out entire U-duality multiplets.  (For example, the general 
extremal black hole solutions in $D=4$ heterotic string and type IIA 
string have been constructed in \cite{cy,ch}.) Of course not all of the
simpler solutions lie in the same multiplet; for example there can be
solutions with different fractions of unbroken supersymmetry, achieved by
using different combinations of $n$-form field strengths, which obviously 
cannot be related by U duality.

     If the dilaton vectors for a set of $N$ field strengths $F_\a$ of rank
$n\ge2$ satisfy the dot products
\be
M_{\a\b}\equiv \vec c_\a \cdot \vec c_\b = 4\delta_{\a\b} -
\fft{2(n-1)(D-n-1)}{D-2}\ ,\label{mmatrix}
\ee
then either they themselves, or a set related to them by the action of the
Weyl group of the U duality group, admit $p$-brane solutions where the
Chern-Simons and $FFA$ terms are not active \cite{lpsol}.  
The maximum value of $N$ depends
on the rank of the field strengths, and on the dimension $D$.  For example, 
for 2-form field strengths, which can be used to construct black holes or
$(D-4)$-branes, $N_{\rm max}=2$ for $6\le D\le 9$;  $N_{\rm max}=3$ in $D=5$;
and $N_{\rm max}=4$ in $3\le D\le 4$.  In fact, we can perform a further 
truncation to the single-scalar Lagrangian
\be
e^{-1} {\cal L}= R -\ft12 (\del\phi)^2 -\fft{1}{2 n!} \, e^{-a\phi}\,
F^2 \label{ss}
\ee
where $a$, $\phi$ and $F$ are given by \cite{lpsol}
\bea
a^2 &=& \Big(\sum_{\a ,\b} (M^{-1})_{\a\b}\Big)^{-1}\ ,\qquad
\phi=a \sum_{\a ,\b} (M^{-1})_{\a\b} \, \vec c_\a\cdot \vec\phi \ ,
\nonumber\\
F_\a^2 &=& a^2 \sum_\b (M^{-1})_{\a\b} \, F^2\ ,\label{cons}
\eea
The parameter $a$ can be conveniently re-expressed as
\be
a^2=\Delta - \fft{2 d\td d}{D-2}\ ,\label{delta}
\ee
where $\Delta=4/N$.   For elementary solutions $d=n-1$, while $d=D-n-1$ 
for solitonic solutions, with $\td d\equiv D-d-2$ in both cases.  All these 
solutions are
supersymmetric, preserving $2^{-N}$ of the supersymmetry for $N\le 3$, and 
$1/8$ for $N=4$.  In these single-scalar $p$-branes, the charges carried by
each field strength $F_\a$ are equal.  They can be generalised to 
multi-scalar solutions where the $N$ charges become independent parameters
\cite{lpmult}.  It was observed in \cite{lpxtoda} that the equations of motion
describing these multi-scalar $p$-branes could be cast into the form of
$N$ Liouville equations, with the constraint that the total Hamiltonian
vanishes.   Note that these simpler solutions, where the Chern-Simons and
$FFA$ terms vanish, can be oxidised to $D=11$ where they can be interpreted
as M-branes, intersecting M-branes [7-12], or boosted 
intersecting M-branes \cite{kklp}.  It should be remarked that there 
also exist many other choices
of sets of field strengths for which the dilaton vectors do not satisfy the
conditions (\ref{mmatrix}) \cite{lpsol}.  For any such set of field strengths, 
it seems that there can be no simple solutions where the  Chern-Simons or 
$FFA$ terms can be neglected, and hence the non-supersymmetric solutions 
described in \cite{lpsol} cannot be embedded in the supergravity theory.

    The situation for 1-form field strengths is a little different.  Their 
potentials are 0-forms, subject to constant shift symmetries. They may be
scalars or pseudo-scalars.
The associated $p$-branes have either $p=-1$ in the elementary case, or
$p=D-3$ in the solitonic case.  The former can be viewed as instantons, and
require that the $D$-dimensional spacetime be Euclideanised to 
positive-definite signature.  There are again consistent truncations possible
when the dilaton vectors of the retained field strengths satisfy 
(\ref{mmatrix}).  In this case, we have $N_{\rm max}=2$ for $7\le D\le 8$;
$N_{\rm max}=4$ for $5\le D\le 6$; $N_{\rm max}=7$ for $D=4$ and 
$N_{\rm max}=8$ for $D=3$.   All of the solutions preserve partial 
supersymmetry. 

     In this paper, we shall show that further consistent truncations are
possible for 1-form field strengths, in certain cases where the dilaton 
vectors do not satisfy (\ref{mmatrix}), namely if the dot products
of the $N$ field strengths instead satisfy the relation
\be
M_{\a\b}= 4 \delta_{\a\b} -2 \delta_{\a,\b +1} -2 \delta_{\a,\b-1} \ .
\label{cartan}
\ee
This is in fact twice the Cartan matrix for $SL(N+1,R)$.  As we shall show, 
this has the consequence that the equations of motion of the 
consistently-truncated system can be cast into the form of the $SL(N+1,R)$
Toda equations, with the Hamiltonian constrained to vanish.  We are thus
able to obtain explicit multi-scalar solutions in these cases.  They can 
also be further
reduced to single-scalar solutions in which the $N$ charges occur in fixed
ratios, determined by (\ref{cons}).  These single-scalar solutions have 
\be
a^2=\Delta=\fft{24}{N(N+1)(N+2)}\ .\label{sunaval}
\ee  

     It is interesting to note that all the $p$-brane solutions for which
the Chern-Simons and $FFA$ terms are not active seem to be associated with
completely-integrable systems of equations.  As we mentioned above, the 
supersymmetric solutions with $\Delta=4/N$ arise as solutions of diagonalised
systems of Liouville equations.  In addition, there is a $\Delta=4$ (
{\it i.e.}\ $a=\sqrt3$) dyonic black hole in $D=4$ \cite{gw,gk}, 
which arises as a solution of the $SL(3,R)$ Toda equations \cite{lpxtoda}.  
This, in common with the
new $SL(N+1,R)$ Toda solutions we shall obtain below, is non-supersymmetric.
All of the non-supersymmetric examples have negative binding energy, in the
sense that they can decay into basic $\Delta=4$ constituents whose total mass
is smaller when widely separated.

\section{Toda $(D-3)$-branes}

      Solitonic $(D-3)$-branes arise as solutions purely of the scalar
and pseudoscalar sector of the supergravity theory.   The bosonic Lagrangian
for this sector in $D$ dimensions can be written in the form
\be
e^{-1} {\cal L} = R - \ft12 (\del \vec \phi)^2 - \ft12 \sum_\a 
e^{- \vec c_\a \cdot \vec \phi}\, F^2_\a\ .\label{genlag}
\ee
In this formulation, the spin-0 fields have been divided into two catgories,
namely dilatonic scalars $\vec\phi = (\phi_1, \phi_2, \ldots)$ which appear
in the exponentials, and the rest, which can be viewed as 0-form potentials
for the 1-form field strengths $F_\a$.  In general the structure of these
field strengths is complicated, owing to the Chern-Simons modifications
coming from dimensional reduction.  In the case of the dimensional reduction
of $D=11$ supergravity, the constant dilaton vectors $\vec c_\a$ that 
characterise the dilaton couplings, and the Chern-Simons modifications to
the field strengths $F_\a$, may be found in \cite{lpsol}. 

      The simplest kind of $(D-3)$-brane is obtained by setting all except
one of the field strengths $F_\a$ to zero, in which case the Lagrangian 
(\ref{genlag}) can be consistently truncated to
\be
e^{-1}{\cal L} = R -\ft12 (\del\phi)^2 - \ft12 e^{-a\phi} F^2\ ,
\label{sslag}
\ee
where $a=2$, and $F=d \chi$ since in this truncation all the Chern-Simons
modifications vanish.  The solution for the $(D-3)$-brane is given by 
\cite{dkl,lpss}
\bea
ds^2 &=& \eta_{\mu\nu} dx^\mu dx^{\nu}  + H^{4/a^2} (dr^2 + r^2d
\theta^2)\ ,\nonumber\\
e^{a\phi/2} &=& H = 1 + k\log r\ ,\qquad \chi = 4Q \theta\ ,
\label{sssol}
\eea
where $Q = k/(2a)$.  The shift symmetry $\chi\rightarrow \chi + 1$ implies
that the charge must be quantised, {\it i.e.}\ $Q=j/(8\pi)$.

     The metric (\ref{sssol}) is not asymptotically flat.  However, the
reason for this is simply that, as usual for supersymmetric $p$-brane 
solutions, the metric and dilaton are given in terms of an harmonic 
function $H$ on the transverse space, and in this case the transverse space 
is two-dimensional, implying a logarithmic harmonic function.  In fact these 
solutions can in general be obtained by the process of vertical dimensional 
reduction from an asymptotically well-behaved $(D-3)$-brane in 
$(D+1)$ dimensions \cite{lpsdr}.  
In this process, one constructs a multi-centered $p$-brane in the higher 
dimension, with a continuum of centers lying along the axis of the extra 
dimension of the three-dimensional transverse space.  As with the analogous
construction of the potential for a uniform line of charge in electrostatics,
the integration over a line of $1/r$ harmonic functions in $R^3$ gives the
$\log r$ harmonic fuction in $R^2$.  The pathologies associated with the
asymptotically-divergent harmonic function can thus be avoided if the 
solution is vertically oxidised to the higher dimension.  In particular, since
the mass per unit spatial $p$-volume is always preserved under dimensional
reduction, one can attach a formal meaning to the mass per unit 
$(D-3)$-volume in $D$ dimensions, despite the absence of an asymptotically
Minkowskian spacetime structure.  For a metric of the form $ds^2= dx^\mu 
dx^\nu \eta_{\mu\nu} + e^{2B}(dr^2 + r^2 d\theta^2)$, the ADM mass per unit 
$(D-3)$-volume is formally given by
\be
m=\ft12 \fft{d B}{d \rho}\Big|_{\rho\rightarrow 0}\ ,\label{mass}
\ee
where $\rho=\log r$.  The anticommutator of supercharges ${\cal Q}_\epsilon$
defines the \bog matrix ${\cal M}= \{{\cal Q}_{\epsilon_1}, 
{\cal Q}_{\epsilon_2}\}=
\epsilon_1^\dagger {\cal M}\epsilon_2$, whose zero eigenvalues correspond
to unbroken components of supersymmetry.  For $p$-brane solutions using the
${\cal F}_1^{(ij)}$ 1-form field strengths, ${\cal M}$ is given by 
\cite{lpsol}
\be
{\cal M} = m \oneone + \ft12 q_{ij}\, \Gamma_{\hat1\hat2 ij} +
\ft12 p_{ij}\, \Gamma_{ij}\ ,\label{bogmat}
\ee
where $q_{ij}$ denote the magnetic charges in the $(D-3)$-brane case, with 
$\hat1, \hat2$ being the transverse-space indices, and $p_{ij}$ denote the 
electric charges in the instanton case. 
For the $(D-3)$-brane given in (\ref{sssol}), just one of the magnetic 
charges, {\it e.g.}\ $q_{12}$, is non-zero, and the eigenvalues $\mu$ of the
\bog matrix ${\cal M}=m\oneone + Q \Gamma_{\hat1\hat2 1 2}$ are given by
$\mu=m\pm Q$.   Since $m=Q$ for this solution, we see that it saturates the 
\bog bound, and that it preserves $\ft12$ of the supersymmetry.  

     Further supersymmetric $(D-3)$-branes arise under certain circumstances
when more than one field strength $F_\a$ carries a charge.  Specifically, if
$N$ field strengths have dilaton vectors $\vec c_\a$ that satisfy the condition
\be
M_{\a\b}\equiv \vec c_\a\cdot \vec c_\b = 4\delta_{\a\b}\ ,\label{mmatrix1}
\ee
then there can be a solution given by \cite{lpmult}
\bea
ds^2 &=& \eta_{\mu\nu} dx^\mu dx^\nu + \Big(\prod_{\a=1}^N H_\a\Big)\,
(dr^2 +r^2 d\theta^2) \ ,\nonumber\\
e^{\varphi_\a} &=& H_\a =1+ k_\a \log r\ , \qquad \chi_\a = 4 Q_\a\, 
\theta \ ,\label{mssol}
\eea
where $Q_\a=\ft14 k_\a$, and $\varphi_\a \equiv \vec c_\a \cdot 
\vec\phi$.
To be precise, not every set of field strengths whose dilaton vectors 
satisfy (\ref{mmatrix}) will give a solution of this form, because the 
Chern-Simons terms will in general contribute.  However, there always
exists some choice of field strengths for which the Chern-Simons terms will
make no contribution.  The full discrete set of choices of field strengths
that satisfy (\ref{mmatrix1}) for a given $N$ forms a multiplet under the
Weyl group of the $U$ duality group \cite{lpsweyl}.  The mass per unit 
$(D-3)$ volume
is equal to the sum of the charges $Q_\a$, and again saturates the Bogomol'nyi
bound.  The solutions preserve varying amounts of supersymmetry depending 
upon the number of $N$ of non-vanishing charges.  For example, they preserve
$\ft14$ for $N=2$, and $\ft18$ for $N=3$.  Further details may be found in
\cite{lpsol}.

       Let us now turn to the construction of the $SL(N+1,R)$ Toda solutions.
These make use of the 1-form field strengths ${\cal F}_1^{(ij)}$ that come 
from the dimensional reduction of the $D=11$ metric.  The full Lagrangian
can be consistently truncated to a sector involving just these, and the
metric and dilatonic scalars:
\be
e^{-1}{\cal L}= R -\ft12 (\del\vec\phi)^2 -\ft12 \sum_{i<j} e^{-\vec b_{ij}
\cdot \vec\phi}\, ({\cal F}_1^{(ij)})^2\ .\label{tlag}
\ee
The dilaton vectors $\vec b_{ij}$ are given in \cite{lpsol}, and satisfy
\be
\vec b_{ij}\cdot \vec b_{k\ell} = 2\delta_{ik} + 2\delta_{j\ell} 
-2 \delta_{i\ell} -2 \delta_{jk} \ .\label{bijdot}
\ee
The full expressions for the field strengths, including Chern-Simons
corrections, are \cite{lpsol}
\be
{\cal F}_1^{(ij)} = \gamma^{kj}\, d {\cal A}_0^{(ik)}\ ,\label{csmod}
\ee
where 
\be
\gamma^{ij}\equiv ((1+{\cal A}_0)^{-1})^{ij}= \delta^{ij} - 
{\cal A}_0^{(ij)} + {\cal A}_0^{(ik)}{\cal A}_0^{(kj)} + \cdots\ .
\label{gamma}
\ee
Note that the 0-form potentials ${\cal A}_0^{(ij)}$, like the 1-form 
field strengths, exist only for $i<j$, and so the series in (\ref{gamma})
terminates after a finite number of terms.  

     It is easy to verify from the above that the field strengths 
${\cal F}_1^{(i,i+1)}$ have no Chern-Simons corrections.  Denoting these
by ${\cal F}_\a \equiv {\cal F}_1^{(\a,\a +1)} =d \chi_\a$, and their 
associated dilaton vectors by $\vec c_\a \equiv \vec b_{\a,\a +1}$, 
which satisfy the dot product relations (\ref{cartan}), we see
that Lagrangian (\ref{tlag}) can be consistently truncated to
\be
e^{-1} {\cal L}= R -\ft12 \sum_{\a,\b=1}^N (M^{-1})_{\a\b}\, 
\del_{\sst M}\varphi_\a 
\del^{\sst M} \varphi_\b-\ft12 \sum_{\a=1}^N e^{-\varphi_\a}\, 
(\del \chi_\a)^2 \ ,\label{ttlag}
\ee
where $\varphi_\a= \vec c_\a\cdot \vec\phi$.  The maximum value of $N$ in
$D$ dimensions is clearly given by $N_{\rm max}=10-D$.  We 
proceed by making the standard metric and field strength ans\"atze
\bea
ds^2 &=& \eta_{\mu\nu} dx^\mu dx^\nu + e^{2B(r)}\, (dr^2 + r^2 d\theta^2 )
\ ,\nonumber\\
\chi_\a &=& 4Q_\a \, \theta\ .\label{ans}
\eea
Substituting into the equations of motion following from (\ref{ttlag}), we
obtain
\bea
&&\varphi_\a'' = -8\sum_\b M_{\a\b}\, Q_\b^2\, e^{-\varphi_\a}\ ,\qquad
B=\sum_{\a,\b} (M^{-1})_{\a\b}\, \varphi_\a\ ,\label{eqs0}\\ 
&&\sum_{\a,\b}(M^{-1})_{\a\b}\, \varphi_\a'\, \varphi_\b' = 16
\sum_\a Q_\a^2\, e^{-\varphi_\a} \ ,\label{eqs1}
\eea
where a prime denotes a derivative with respect to $\rho=\log r$.  
Making the redefinition  $\Phi_\a =
-2 \sum_\b (M^{-1})_{\a\b}\, \varphi_\b$, these equations become
\bea
&& \Phi_\a'' = 16 Q_\a^2\, \exp(\ft12 \sum_\b M_{\a\b}\, \Phi_\b)\ ,
\qquad B=-\ft12 \sum_\a \Phi_\a\ ,\nonumber\\
&& \sum_{\a,\b} M_{\a\b} \Phi_\a'\, \Phi_\b' = 64 \sum_\a Q_\a^2 \, 
\exp(\ft12 \sum_\b M_{\a\b}\, \Phi_\b)\ .
\eea
The further redefinition $\Phi_\a=q_\a -4 \sum_\b (M^{-1})_{\a\b} 
\log(4Q_\b)$ removes the charges from the equations, giving
\bea
q_1''&=& e^{2q_1 -q_2}\ ,\nonumber\\
q_2''&=& e^{-q_1+2q_2-q_3}\ ,\nonumber\\
q_3''&=& e^{-q_2 + 2q_3 -q_4}\ ,\label{suntoda}\\
&&\cdots \nonumber\\
q_{\sst N}'' &=& e^{-q_{\sst{N}-1} + 2q_{\sst N}}\ .\nonumber
\eea
These are precisely the $SL(N+1,R)$ Toda equations.  The solution is subject
to the further constraint (\ref{eqs1}), which, in terms of the $q_\a$, 
becomes the constraint that the Hamiltonian
\be
{\cal H} = 4\sum_{\a,\b} (M^{-1})_{\a\b}\, p_\a\, p_\b -
           \sum_\a \exp(\ft12 \sum_\b M_{\a\b}\, q_\b) \label{hamil}
\ee
for the Toda system (\ref{suntoda}) vanishes.   

     The general solution to the $SL(N+1,R)$ Toda equations is presented
in an elegant form in \cite{a}:
\be
e^{-q_\a}= \sum_{k_1<\cdots < k_\a} f_{k_1}\cdots f_{k_\a} \,
\Delta^2(k_1,\ldots,k_\a)\, e^{(\mu_{k_1}+\cdots \mu_{k_\a})\rho}\ ,
\label{gensol}
\ee
where $\Delta^2(k_1,\ldots,k_\a)=\prod_{k_i <k_j} (\mu_{k_i}- 
\mu_{k_j})^2$ is the Vandermonde determinant, and $f_k$ and $\mu_k$ are
arbitrary constants satisfying
\be
\prod_{k=1}^{N+1} f_k = - \Delta^{-2}(1,2,\ldots,N+1)\ ,\qquad
\sum_{k=1}^{N+1} \mu_k =0\ .
\ee
The Hamiltonian, which is conserved, takes the value ${\cal H}=\ft12 
\sum_{k=1}^{N+1} \mu_k^2$.

     The solution (\ref{gensol}) in general involves exponential  
functions of $\rho$.  Furthermore, the vanishing of the Hamiltonian implies
that the parameters $\mu_k$, and hence the solutions, will in general be
complex.  However, there exists a limit, under which all the $\mu_k$ 
constants vanish, which achieves a vanishing Hamiltonian and real solutions
that are finite polynomials in $\rho$.  Since we are constructing 
$(D-3)$-branes in $D\ge 3$, it follows that we are interested in obtaining
solutions to the $SL(N+1,R)$ Toda equations for $N\le 7$.  When $N=1$, the
Toda system reduces to the Liouville equation, giving rise to the usual
single field strength solution that preserves $1/2$ the supersymmetry, namely
\be
e^{-q_1}=1 + 4 Q \,\rho \ . \label{liousol}
\ee
Note that since there is only a single independent $\mu$ 
parameter when $N=1$, which has to be zero by the Hamiltonian constraint, 
(\ref{liousol}) is in fact the only solution in this case.
   
     For $N=2$, we find that the polynomial solution to the $SL(3,R)$ Toda 
equations (\ref{suntoda}) is
\bea
e^{-q_1} &=& a_0 + a_1\, \rho + \ft12 \, \rho^2\ ,\nonumber\\
e^{-q_2} &=& a_1^2 - a_0 + a_1\, \rho + \ft12 \, \rho^2\ ,
\eea
where $a_0$ and $a_1$ are constants that are related to the charge parameters
$Q_1$ and $Q_2$, on using the boundary condition that the dilatonic scalars,
and hence $\Phi_\a$, vanish ``asymptotically'' ({\it i.e.\ }at $\rho=0$).  
Thus we have 
\be
a_0= \ft{1}{16} Q_1^{-4/3}\, Q_2^{-2/3}\ , 
\qquad a_1=\ft14  Q_1^{-2/3}\,  Q_2^{-2/3}\,
( Q_1^{2/3}+ Q_2^{2/3})^{1/2} \ ,
\ee
which implies that the metric is
\be
ds^2 = \eta_{\mu\nu}dx^\mu dx^\nu + T_1 T_2 (dr^2 + r^2 d\theta^2)
\ ,\label{su3metric}
\ee
where 
\bea
T_1 &=& 1+ 4 Q_1^{2/3}\, ( Q_1^{2/3}+ Q_2^{2/3})^{1/2}\, \rho
+ 8 Q_1^{4/3}\, Q_2^{2/3}\, \rho^2\ ,\nonumber\\
T_2 &=&1+ 4 Q_2^{2/3}\, ( Q_1^{2/3}+ Q_2^{2/3})^{1/2}\, \rho
+ 8 Q_2^{4/3}\, Q_1^{2/3}\, \rho^2\ ,\label{ts}
\eea
and $\rho=\log r$.
It follows from (\ref{mass}) that the mass per unit $(D-3)$-volume is given by
\be
m= ( Q_1^{2/3}+ Q_2^{2/3})^{3/2}\ .\label{su3mass}
\ee
This rather unusual looking mass formula in fact also arises in the 
$a=\sqrt3$ four-dimensional dyonic black hole \cite{gk}.  In that case also,
the equations of motion can be cast into the form of the $SL(3,R)$ Toda
equations \cite{lpxtoda}.  For non-vanishing Hamiltonian the black hole
is non-extremal, becoming extremal when the Hamiltonian vanishes. 
The mass formula (\ref{su3mass}) implies that the solution
describes a system with negative binding energy, since the total mass of the
widely-separated constituents is given by $m_\infty=Q_1+Q_2$, which is 
smaller than $m$.  The \bog
matrix in this case is ${\cal M}=m\oneone + Q_1 \Gamma_{\hat1\hat2 1 2} +
Q_2 \Gamma_{\hat1\hat2 2 3}$, and therefore its eigenvalues are
\be
\mu= m \pm \sqrt{Q_1^2 + Q_2^2}\ .
\ee
It follows from (\ref{su3mass}) that the $\mu$ is strictly positive, and
hence the \bog bound is exceeded and there is no supersymmetry, unless either
$Q_1$ or $Q_2$ vanishes.

    For $N=3$, we find the following polynomial solution of the $SL(4,R)$
Toda equations:
\bea
e^{-q_1}&=& a_0 + a_1\, \rho + a_2\, \rho^2 + \ft16 \rho^3\ ,\nonumber\\
e^{-q_2}&=& a_1^2 -2 a_0 a_2 + (2 a_1 a_2 -a_0)\, \rho + 
2 a_2^2 \, \rho^2 + \ft23 a_2\,  \rho^3\  + \ft1{12} \rho^4\ ,\label{su4sol}\\
e^{-q_3}&=& a_0-4 a_1 a_2 + 8a_2^3  +(4 a_2^2 - a_1)\, \rho + a_2\, \rho^2 
+ \ft16 \rho^3\ ,\nonumber
\eea
where the constants $a_0$, $a_1$ and $a_2$ are determined in terms of the 
charges $Q_1$, $Q_2$ and $Q_3$ by the requirement that the dilatonic scalars
vanish at $\rho=0$.  This implies that
\be
e^{q_\a(0)} = \prod_\b (4 Q_\b)^{4 (M^{-1})_{\a\b}}\ ,\label{sunconst}
\ee
and hence
\bea
&&a_0= \fft1{64} Q_1^{-3/2} \, Q_2^{-1}\, Q_3^{-1/2}\ ,\qquad
a_1^2 -2 a_0 a_2= \fft1{256} Q_1^{-1}\, Q_2^{-2}\, Q_3^{-1}\ ,\nonumber\\
&&a_0-4a_1 a_2 +8a_2^3 =\fft1{64} Q_1^{-1/2} \, Q_2^{-1}\, Q_3^{-3/2}\ .
\label{3charge}
\eea
The metric is given by (\ref{ans}), with
\be
e^{2B}=\prod_\a e^{q_\a(0)-q_\a} \ ,\label{sunmetric}
\ee
and hence
\bea
m=\fft{a_1}{4 a_0} + \fft{2a_1 a_2-a_0}{4(a_1^2-2a_0 a_2)} + 
\fft{4a_2^2 -a_1}{4(a_0-4 a_1 a_2 + 8a_2^3)}\ .\label{3mass}
\eea
Thus we find that the mass is given in terms of the charges by the positive 
root of the sextic
\bea
&&m^6- (3Q_1^2 + 2 Q_1 Q_3 + 3 Q_3^2 + 3Q_2^2)m^4 -36 \sqrt{Q_1 Q_3}Q_2
(Q_1 +Q_3) m^3 \nonumber \\
&&+\Big[(Q_1+Q_3)^2 (3Q_1^2 -2 Q_1 Q_3 + 3Q_3^2) - Q_2^2 (21 Q_1^2 + 122 
Q_1 Q_3 + 21 Q_3^2) + 3 Q_2^4 \Big] m^2 \nonumber \\
&&+ 4\sqrt{Q_1 Q_3} Q_2 (Q_1+Q_3) 
(9 Q_1^2 -14 Q_1 Q_3 + 9Q_3^2 -18Q_2^2) m\label{dgsbrkfst}\\
&&-(Q_1-Q_3)^2(Q_1+Q_3)^4 -Q_2^2(3Q_1^4-68 Q_1^3 Q_3 +114 Q_1^2 Q_3^2 -
68 Q_1 Q_3^2 + 3Q_3^4)\nonumber\\
&& -Q_2^4(3Q_1^2 + 38 Q_1 Q_3 + 3Q_3^2) -Q_2^6 =0\ .\nonumber
\eea
There seems to be  no way to give an explicit closed-form expression for 
the mass in terms of
the charges.  The \bog matrix ${\cal M}= m\oneone + Q_1 \Gamma_{\hat1\hat2 12}
+ Q_2\Gamma_{\hat1\hat2 2 3} + Q_3 \Gamma_{\hat1 \hat2 3 4}$ has eigenvalues
\be
\mu=m\pm \sqrt{(Q_1\pm Q_3)^2 + Q_2^2} \ ,
\ee
where the two $\pm$ signs are independent.  For generic values of the charges,
$\mu>0$ and the solution has no supersymmetry.  If $Q_2=0$, the solution
reduces to the two-charge supersymmetric solution, preserving $\ft14$ of the
supersymmetry.  In this case, the $SL(4,R)$ Toda equations reduce to two
decoupled Liouville equations.

     For higher values of $N$, the explicit forms of the polynomial solutions
to the $SL(N+1,R)$ Toda equations become increasingly complicated.  The 
structure of these polynomials can be summarised as follows.  For each $N$, 
we find that $e^{-q_\a}$ are polynomials in $\rho$ of degree
$n_\a=\a(N+1-\a)$, {\it i.e.}
\be
\fft{d^{n_\a+1}}{d\rho^{n_\a+1}} e^{-q_\a}= 0\ .
\ee
After substituting these into the $SL(N+1,R)$ Toda equations (\ref{suntoda}),
we find that there are $N$ independent parameters, which can be related to
the $N$ charges $Q_\a$ by equation (\ref{sunconst}).  The metric is given by 
(\ref{ans}) with $e^{2B}$ again given by (\ref{sunmetric}).  The mass is given
in terms of charges by an $N!$'th-order polynomial equation.  Although it
appears not to be possible to give closed-form expressions for the mass in
terms of the charges for $N\ge3$, we expect nevertheless that it is less than
the sum of the charges, indicating again that they are bound states with
negative binding energies.  One can see this explicitly in the special case
where the charges have the fixed ratio given by
\be
Q_\a = a Q \Big(\sum_\b (M^{-1})_{\a\b} \Big)^{1/2}
=\ft12 a Q \sqrt{\a(N+1-\a)} \ ,\label{suncharge}
\ee
where $a$ is given by (\ref{sunaval}).  Under these circumstances the
solutions reduce to single-scalar solutions, given by (\ref{sssol}), and
have mass
\be
m=\fft{2Q}{a}\ .
\ee
It is easy to verify that this is always larger than the total mass of
the widely-separated constituents, $m_\infty=\sum_\a Q_\a$.  The calculation
of the eigenvalues of the \bog matrix becomes increasingly complicated with
increasing $N$.  For example, for the $SL(5,R)$ case we find
\be
\mu=m \pm \sqrt{Q_1^2 + Q_2^2 + Q_3^2 + Q_4^2 \pm 2
\sqrt{(Q_1 Q_3)^2 + (Q_1 Q_4)^2 + (Q_2 Q_4)^2}}\ ,
\ee
whilst for $SL(6,R)$ we find that $\mu=m\pm \kappa$, where $\kappa$ denotes
the roots of the quartic equation $\kappa^4 -2 \kappa^2\, \a -8 \kappa \, Q_1
\, Q_3 \, Q_5 + \b=0$, and 
\bea
\a&=&Q_1^2 + Q_2^2 + Q_3^2 + Q_4^2 + Q_5^2\ ,\\
\b&=& a^2 -4\Big( (Q_1 Q_3)^2 + (Q_1 Q_4)^2 +  (Q_1 Q_5)^2 +
          (Q_2 Q_4)^2 +  (Q_2 Q_5)^2 +  (Q_3 Q_5)^2 \Big)\ .\nonumber
\eea
For all $N$, the solutions are non-supersymmetric for generic values of the
charges.  However, they can be reduced to the previously-known supersymmetric
solutions if appropriate charges are set to zero, such that the remaining
charges $Q_\a$ have non-adjacent indices.  In these cases, the solutions
preserve $2^{-n}$ of the supersymmetry, where $n$ is the number of charges
remaining.

\section{Toda Instantons}

     In the previous section we discussed $(D-3)$-branes solutions, where 
the 1-form field strengths carry magnetic charges.  We can also discuss
the case where the 1-forms instead carry electric charges.  In this case, the
solutions will describe $p$-branes with $p=-1$, which can be interpreted as
instantons.  Since there is no longer a time direction, it is necessary first
to Euclideanise the supergravity theories, by performing a Wick rotation of
the time coordinate.  At the same time, account must be taken of the parities
of the various fields in the theory, since the kinetic terms for fields of
odd parity will undergo a sign change.  This has been
discussed in detail for the Type IIB superstring in \cite{ggp}, where it was
argued that the kinetic term for the R-R scalar $\chi$ undergoes such a
reversal of sign.  In our case, where we keep the subset of 1-forms
${\cal F}_1^{(12)}$, ${\cal F}_1^{(23)}$, ${\cal F}_1^{(34)},\cdots$ in 
$D\le 9$ dimensions, the potential for ${\cal F}_1^{(12)}$ is precisely
this same field $\chi$ if we follow the type IIB reduction route.  By the
same token, we may argue that the potentials for the other 1-forms should 
also be viewed as having odd parity.  As a check, one may verify that this
assignment is consistent with the structure of the Chern-Simons modifications
for the entire set of field strengths in the supergravity theory.  Thus we 
are led to replace the consistently-truncated Lagrangian (\ref{ttlag}) by
\be
e^{-1} {\cal L}= R -\ft12 \sum_{\a,\b=1}^N (M^{-1})_{\a\b}\,
\del_{\sst M}\varphi_\a
\del^{\sst M} \varphi_\b + \ft12 \sum_{\a=1}^N e^{-\varphi_\a}\,
(\del \chi_\a)^2 \ \label{eucttlag}
\ee
in the Euclideanised theory.

     In the instanton solutions, the metric is flat, and can be written as
\be
ds^2= dr^2 + r^2 d \Omega^2 \ ,\label{flatmetric}
\ee
where $d\Omega^2$ is the metric on the unit $(D-1)$-sphere.  The standard
elementary ansatz for the 1-form field strengths becomes simply
\be
F_\a = d\chi_\a = 4 Q_\a\, e^{\varphi_\a}\, dr \ ,\label{elans}
\ee
and the remaining equations of motion become
\be
\varphi_\a'' = \fft{8}{(D-2)^2}\sum_\b Q_\b^2\,  e^{\varphi_\b}\ ,
\label{eucsun}
\ee
where a prime denotes the derivative with respect to $\rho
\equiv r^{2-{\sst D}}$.  Following analogous steps to those 
described in the previous
section, we find that (\ref{eucsun}) reduces to the $SL(N+1,R)$ Toda equations
(\ref{suntoda}), where
\be
\varphi_\a = \ft12\sum_\b M_{\a\b}\, q_\b -
2 \log\Big( \fft{4 Q_\a}{D-2}\Big) \ .
\ee
We now obtain the desired instanton solutions by taking the same solutions
of the Toda equations that we discussed in the previous section, namely 
those that are finite polynomials in $\rho$.  It is worth remarking, 
however, that since $\rho$ is now equal to $r^{2-{\sst D}}$ rather than 
to $\log r$, the instanton solutions are well-behaved asymptotically at large
$r$.  

\section{$SL(3,R)$ Toda $p$-branes in the type IIB string}

        Having obtained the non-supersymmetric $SL(N+1, R)$ Toda 
$(D-3)$-branes and instanton solutions in $3 \le D \le 8$ dimensions, it is
of interest to investigate how these solutions be interpreted in M-theory in
$D=11$ or string theory in $D=10$. All these Toda $p$-brane solutions involve
1-form field strengths ${\cal F}^{(\a, \a+1)}$ which arise from the dimensional
reduction of the metric in $D=11$.  Thus the corresponding 11-dimensional
oxidations of these solutions involve a twisted metric.  On the other hand,
from the type IIB string point of view, the field strength ${\cal F}^{(12)}$ 
is the dimensional reduction of the derivative of the R-R scalar $\chi$, and
${\cal F}^{(23)}$ is the dimensional reduction of the NS-NS 3-form field 
strength.   Thus the Toda solutions that involve only ${\cal F}^{(12)}$ and
${\cal F}^{(23)}$ can be oxidised into type IIB solutions in $D=10$ whose
metrics will be diagonal.  For example the two-charge $5$-brane in $D=8$ can
be oxidised to a 5-brane intersecting a 7-brane in the $D=10$ type IIB theory.
The metric is given by
\be
ds^2 = T_2^{-1/4} dx^\mu dx^\nu \eta_{\mu\nu} + T_2^{3/4} T_1 (dr^2 +
r^2 d\theta^2) + T_2^{3/4} (dz_1^2 + dz_2^2) \ ,\label{intersect}
\ee
where $T_1$ and $T_2$ are given by (\ref{ts}) with $\rho = \log r$.  This 
describes a 7-brane if $Q_2=0$; if instead $Q_1=0$, it describes a planar
continuum of 5-branes.  The interpolation when both $Q_1$ and $Q_2$ are 
non-zero provides the interpretation as an intersection of a 5-brane and
a 7-brane.

      On the other hand, the two-charge instanton solution in $D=8$ can be 
oxidised to a string intersecting an instanton in $D=10$ type IIB theory with 
Euclidean signature.  The metric is 
\be
ds^2 = T_2^{-3/4} (dz_1^2 + dz_2^2) + T_2^{1/4}(dr^2 + r^2 d\Omega^2)
\ ,\label{string}
\ee
and $e^\phi= T_1\, T_2^{-1/2}$, 
where $T_1$ and $T_2$ are given by (\ref{ts}) with $\rho = r^{-6}$.  
When $Q_1=0$, it is
nothing but the NS-NS string of type IIB theory in $D=10$ Euclidean space.
If instead $Q_2=0$, the space is flat because of the cancellation of the 
enery-momentum tensors between the dilaton and the R-R scalar.  It describes
a planar continuum of supersymmetric instantons, of the type obtained in 
type IIB supergravity in \cite{ggp}.

     It is worth remarking that the NS-NS string solution in (\ref{string})
carries a real charge, if the NS-NS 3-form has odd parity. In fact U duality
of the type IIB theory in Euclidean space requires that if the R-R scalar 
has odd parity, then one
of the 3-forms must have odd parity and the other must have even parity.  Thus
U duality interchanges real and imaginary 3-form charges in the Euclidean
type IIB string, which is reminiscent of the electromagnetic duality in
the Euclidean $D=4$ Maxwell equations \cite{hr}.  With the opposite choice 
of parity assignments for the 3-forms, the string in (\ref{string}) would
carry imaginary electric charge, which would be analogous to the Euclidean
black holes with imaginary electric charge discussed in \cite{hr}.
It is not altogether clear
in the supergravity case, however, what the proper parity assignments
should be.  For example, the ${\cal F}_1^{(12)}$ field strength was argued
to have odd parity in the type IIB theory \cite{ggp}. However, this same 
field strength, reduced to $D=9$, can also be obtained from the dimensional 
reduction of the metric in $D=11$ Euclidean supergravity, which seems to 
suggest that it should instead have even parity.  Possibly any assignment of 
parities that is compatible with U duality is permissable.

\section{Conclusions and discussion}

     The bosonic Lagrangian for maximal supergravity dimensionally reduced
to $D$ dimensions is in general rather complicated, owing to the occurrence
of Chern-Simons modifications to many of the field strengths.  In certain
rather special circumstances, $p$-brane solutions can be found that 
correspond to consistent truncations of the Lagrangian in which only
field strengths without Chern-Simons modifications are present.   Solutions 
using such subsets of the fields are much simpler in form than the more
generic ones, and therefore are also much easier to obtain.  Most of the
known solutions of this simpler form are supersymmetric.  The equations of
motion that describe them can be recast in the form of one or more decoupled
Liouville equations.  A non-supersymmetric example was also already known,
namely an $a=\sqrt3$ black-hole dyon in $D=4$ \cite{gw,gk}.  The equations of
motion in this case can be cast into the form of the $SL(3,R)$ Toda equations 
\cite{lpxtoda}.  In this paper, we have investigated a further class of 
non-supersymmetric solutions, in which certain of the 1-form field strengths
carry magnetic or electric charges, giving rise to $(D-3)$-branes or
instantons respectively.  The equations of motion again turn out to be
those of completely integrable systems, namely the $SL(N+1,R)$ Toda equations.
These solutions reduce to the previously-known multiply-charged 
supersymmetric solutions when appropriate charges vanish.

     In most of the paper, we concentrated on describing the $D$-dimensional
$SL(N+1,R)$ Toda $p$-branes in terms of the Kaluza-Klein reduction of
M-theory.  We showed also that there is an $SL(3,R)$ Toda solution in the
type IIB theory in $D=10$, which is obtained by oxidising the $SL(3,R)$ 
Toda 5-brane in $D=8$ {\it via}\ the alternative type IIB pathway.  The
$D=10$ solution has the interpretation of an intersection between a 7-brane
and a 5-brane.  In particular, when the charge associated with the 5-brane 
vanishes, it gives rise to a supersymmetric 7-brane in the type IIB theory.
In this case, the function $T_2$ in (\ref{intersect}) becomes harmonic.
If it is now taken to depend on only one of the two transverse-space Cartesian
coordinates $y_1=r \cos\theta$ or $y_2=r\sin\theta$, then one obtains the
7-brane solution discussed in \cite{brgpt}, where $T_2$ is a linear 
function of
the chosen coordinate.  On the other hand $\chi$ is proportional to the other
transverse coordinate, which must therefore be periodic because of the 
$\chi\rightarrow \chi+1$ shift symmetry of the R-R scalar. (Compactifying the
type IIB theory on this periodic coordinate gives rise to massive $N=2$
supergravity in $D=9$ \cite{brgpt}.)  Both of the above 7-branes in the 
type IIB theory should be distinguished from the modular-invariant 
7-brane \cite{ggp}, which is the oxidation of the cosmic string in $D=4$ 
\cite{gsvy}. 

\section*{Acknowledgement}

    We are grateful to G.W. Gibbons, K.S. Stelle and P.K. Townsend for
useful discussions.

\vfill\eject

\end{document}